\documentclass[english,a4paper,preprint]{revtex4}
\usepackage[T1]{fontenc}
\usepackage[latin1]{inputenc}

\makeatletter
\@ifundefined{textcolor}{}
{%
 \definecolor{BLACK}{gray}{0}
 \definecolor{WHITE}{gray}{1}
 \definecolor{RED}{rgb}{1,0,0}
 \definecolor{GREEN}{rgb}{0,1,0}
 \definecolor{BLUE}{rgb}{0,0,1}
 \definecolor{CYAN}{cmyk}{1,0,0,0}
 \definecolor{MAGENTA}{cmyk}{0,1,0,0}
 \definecolor{YELLOW}{cmyk}{0,0,1,0}
 }

\makeatletter

\makeatletter

\makeatletter

\usepackage{geometry}

\makeatother

\makeatother

\usepackage{babel}

\makeatother

\usepackage{babel}

\makeatother

\usepackage{babel}

\begin{document}

\title{Euler equations in a 3+1 framework}

\author{A. R. Sagaceta-Mejia{*}, A. L. Garcia-Perciante{*}}

\address{{*}Depto. de Matematicas Aplicadas y Sistemas, Universidad Autonoma
Metropolitana-Cuajimalpa, Artificios 40 Mexico D.F 01120, Mexico.}
\begin{abstract}
In this paper we show how the non-relativistic transport equations
for a simple fluid can be obtained using a 3+1 representation. A pseudo-galilean
transformation is introduced in order to obtain the Euler conservation
laws. The interpretation of such transformation as well as the implications
and potential extensions of the formalism are briefly discussed. 
\end{abstract}
\maketitle

\section{Introduction}

Kinetic theory provides a fundamental framework to obtain transport
equation for fluids through the Enskog equation. Usually, in the foundations
of irreversible thermodynamics \cite{key-1,key-2}, these equations
are obtained by considering a Maxwell-Boltzmann distribution as the
local equilibrium function and using the fact that the relation between
the hydrodynamic, molecular and chaotic velocities -$\vec{u}$, $\vec{v}$
and $\vec{k}$ respectively- is given by $\vec{v}=\vec{k}+\vec{u}$.

In this work we propose an alternative method to obtain the Euler
equations for a simple non-relativistic gas. This scheme is based
on the approach proposed by A. Sandoval and L. S. Garcia-Colin in
Ref.\cite{key-3} for the relativistic gas. The formalism we develop
in this paper consists in working the non-relativistic case in a similar
way as the relativistic fluid equations are obtained in special relativity
\cite{key-4.1}. We formulate the Enskog transport equation as a conservation
law for four-fluxes, namely the particle flux and the energy-momentum
tensor. In this scheme one should be able to obtain transport equations
to any order in the gradients by calculating the fluxes in the comoving
frame and using an apropriate transformation law.

In Section II we construct the 3+1 scheme and obtain the conservation
laws for a fluid at rest. In Section III we propose a transformation
law for the four-fluxes and obtain the transport equations using the
Enskog equation for the transformed quantites. Finally, a discussion
and final remarks are included in Section IV.

\section{Fluxes in the comoving frame and conservation laws}

In kinetic theory, the conservation laws, given by the Enskog transport
equation, are obtained from the Boltzmann equation which, for a simple
fluid, reads\begin{equation}
\frac{df}{dt}=J\left(ff'\right)\label{eq:1}\end{equation}
where $f$, the distribution function, gives the probability of finding
a molecule in a given cell of phase space. The term on the right side
is the collision kernel whose details are irrelevant for the purpose
of this work. From Eq. (\ref{eq:1}) one can readily obtain the trasport
equation \cite{key-1,key-2}

\begin{equation}
\frac{\partial}{\partial t}\left[n\left\langle \psi\right\rangle \right]+\nabla\cdot\left[n\left\langle \vec{v}\psi\right\rangle \right]=0\label{eq:2}\end{equation}
 where $\psi$ is a collision invariant and $n$ the number density,
given by \begin{equation}
n=\int f\, d^{3}v\label{eq:3}\end{equation}
Thus, for $\psi=1$ the continuity equation is obtained and for $\psi=m\vec{v}$
and $\psi=mv^{2}/2$ the momentum and internal energy balances respectively.

As in the relativistic case, we propose now a 3+1 representation by
considering a fourth (time) component of the position vector. That
is, if\begin{equation}
x^{\mu}=\left(\begin{array}{c}
x\\
y\\
z\\
ct\end{array}\right)\,,\end{equation}

\begin{equation}
v^{\mu}=\left(\begin{array}{c}
v_{x}\\
v_{y}\\
v_{z}\\
c\end{array}\right)\,,\end{equation}
, Eq. (\ref{eq:2}) can be written as\begin{equation}
\left(n\left\langle v^{\mu}\psi\right\rangle \right)_{;\mu}=0\label{eq:4}\end{equation}
Here and throughout this work a semicolon represents a covariant derivative,
latin indices run from 1 to 3, greek ones from 1 to 4 and the usual
summation convension is used. Next, we define the particle and energy-momentum
four-fluxes as\begin{equation}
N^{\alpha}=n\left\langle v^{\alpha}\right\rangle \label{eq:5}\end{equation}
 and \begin{equation}
T^{\ell\nu}=n\left\langle v^{\ell}v^{\nu}\right\rangle \label{eq:6}\end{equation}
 respectively. Notice that $T^{44}$ has been left unspecified. With
these definitions, Eq. (\ref{eq:4}) is a statement of conservation
of these quantities and can be written as\begin{equation}
N_{;\alpha}^{\alpha}=0\label{eq:7}\end{equation}
 and\begin{equation}
T_{;\nu}^{\mu\nu}=0\label{eq:8}\end{equation}
Equations (\ref{eq:7}) and (\ref{eq:8}) yield the particle number
density and momentum conservation laws. However, the energy balance
must be introduced apart, if one considers $T^{4\nu}=n\left\langle v^{4}v^{\nu}\right\rangle $
since in that case $T_{;\nu}^{4\nu}=0$ only yields an identity. In
order to obtain an energy balance equation from the conservation of
an energy-momentum tensor, and following the ideas of the 3+1 formalism
in special relativity, we propose instead \begin{equation}
T^{\mu\nu}=n\left(\begin{array}{cccc}
\left\langle v^{1}v^{1}\right\rangle  & \left\langle v^{2}v^{1}\right\rangle  & \left\langle v^{3}v^{1}\right\rangle  & \left\langle v^{4}v^{1}\right\rangle \\
\left\langle v^{1}v^{2}\right\rangle  & \left\langle v^{2}v^{2}\right\rangle  & \left\langle v^{3}v^{2}\right\rangle  & \left\langle v^{4}v^{2}\right\rangle \\
\left\langle v^{1}v^{3}\right\rangle  & \left\langle v^{2}v^{3}\right\rangle  & \left\langle v^{3}v^{3}\right\rangle  & \left\langle v^{4}v^{3}\right\rangle \\
\left\langle v^{1}v^{4}\right\rangle  & \left\langle v^{2}v^{4}\right\rangle  & \left\langle v^{3}v^{4}\right\rangle  & E\end{array}\right)\,,\label{eq:9}\end{equation}
where $E$ is the total energy per particle, including a term corresponding
to the rest energy. With these definitions one obtains, for the simple
fluid ate rest\begin{equation}
N^{\alpha}=\left(\begin{array}{c}
0\\
0\\
0\\
nc\end{array}\right)\,,\label{eq:10}\end{equation}

\begin{equation}
T^{\mu\nu}=\left(\begin{array}{cccc}
p & 0 & 0 & 0\\
0 & p & 0 & 0\\
0 & 0 & p & 0\\
0 & 0 & 0 & \varepsilon\end{array}\right)\,,\label{eq:11}\end{equation}
where $p=nkT$ is the hydrostatic pressure and $\varepsilon$ includes
both internal energy and rest energy sich that, for the ideal gas
\mbox{$\varepsilon=mc^{2}+\frac{3}{2}kT$}. From these four-fluxes
one can directly obtain the conservation equations. This calculation
is carried out in the labratory frame in the next section where Euler's
equations are obtained.

\section{A pseudo-galilean transformation}

In this section we consider the case of a fluid with hydrodynamic
velocity

\begin{equation}
u_{\nu}=\left(\begin{array}{c}
u_{1}\\
u_{2}\\
u_{3}\\
c\end{array}\right)\label{eq:12}\end{equation}
for which the conservation equations will be obtained by calculating
Eqs. (\ref{eq:7}) and (\ref{eq:8}) for transformed fluxes. In order
to obtain those quantities, we propose the use of a pseudo-galilean
transformation which is the non-relativistic limit of the Lorentz
transformation, namely

\begin{equation}
\mathcal{G}^{\mu\nu}=\left(\begin{array}{cccc}
1 & 0 & 0 & \frac{u_{x}}{c}\\
0 & 1 & 0 & \frac{u_{y}}{c}\\
0 & 0 & 1 & \frac{u_{z}}{c}\\
\frac{u_{x}}{c} & \frac{u_{y}}{c} & \frac{u_{z}}{c} & 1+\frac{u^{2}}{2c^{2}}\end{array}\right)\label{eq:13}\end{equation}
Notice that, for $\mathcal{G}^{44}$ we suggest keeping second order
in $u/c$. This is necessary since the energy equation is quadratic
in the velocities. Moreover, as will be shown in the next section,
the zeroth order terms only account for the thermal part of the total
energy while the second order terms include the mechanical effects.
The implications of including this term will be discussed to some
detail in the last section of this work. 

As in the relativistic case \cite{key-3,key-4.1}, the transformed
quantities are\[
\bar{N}^{\mu}=\mathcal{G}_{\nu}^{\mu}N^{\nu}\]
\[
\bar{T}^{\mu\nu}=\mathcal{G}_{\alpha}^{\mu}\mathcal{G}_{\beta}^{\nu}T^{\alpha\beta}\]

\noindent Using the transformation given by Eq. (\ref{eq:13}) one
obtains\begin{equation}
\bar{N}^{\ell}=nu^{\ell}\,,\label{eq:14}\end{equation}

\begin{equation}
\bar{N}^{4}=nc\left(1+\frac{u^{2}}{2c^{2}}\right)\,,\end{equation}

\begin{equation}
\bar{T}^{k\ell}=p\delta^{k\ell}+\frac{\varepsilon}{c^{2}}u^{k}u^{\ell}\,,\label{eq:15-a}\end{equation}

\begin{equation}
\bar{T}^{4\ell}=\left[n\varepsilon\left(1+\frac{u^{2}}{2c^{2}}\right)+p\right]\frac{u^{\ell}}{c}\,,\label{eq:15}\end{equation}

\begin{equation}
\bar{T}^{44}=n\varepsilon\left(1+\frac{u^{2}}{2c^{2}}\right)+p\frac{u^{2}}{c^{2}}\,.\label{eq:16}\end{equation}

\noindent Calculating the four-divergence of $\bar{N}^{\alpha}$ one
obtains \begin{equation}
\bar{N}_{,\alpha}^{\alpha}=\left(nu^{\ell}\right)_{,\ell}+\frac{\partial n}{\partial t}+n\frac{u}{2c}\frac{\partial u}{\partial t}=0\,,\label{eq:17}\end{equation}

\noindent where, neglecting the last term, using $u<<c$, the usual
continuity equation is obtained.

\noindent The momentum balance equation is obtained from $\bar{T}_{,\nu}^{\ell\nu}=0$,
namely

\begin{eqnarray}
\left(p\delta^{k\ell}+\frac{n\varepsilon}{c^{2}}u^{k}u^{\ell}\right)_{,\ell}+\phantom{jklasjfdlafdhjdjf}\nonumber \\
+\frac{1}{c}\frac{\partial}{\partial t}\left\{ \left[n\varepsilon\left(1+\frac{u^{2}}{2c^{2}}\right)+p\right]\frac{u^{k}}{c}\right\} =0\,.\label{eq:18}\end{eqnarray}

\noindent Since the fluid here considered is non-relativistic, the
internal energy term is \mbox{$\frac{\varepsilon}{c^{2}}=m\left(1+\frac{3}{2}\frac{kT}{mc^{2}}\right)\sim m$}
such that Eq. (\ref{eq:18}) is precisely the momentum balance \begin{equation}
\frac{\partial}{\partial t}\left(nu^{k}\right)+\left(p\delta^{k\ell}+nmu^{k}u^{\ell}\right)_{,\ell}=0\,.\label{eq:19}\end{equation}

\noindent Finally, the energy balance is obtained from $T_{,\nu}^{4\nu}=0$,

\begin{eqnarray}
\left\{ \left[n\varepsilon\left(1+\frac{u^{2}}{2c^{2}}\right)+p\right]u^{\ell}\right\} _{,\ell}+\phantom{jklasjfdl}\nonumber \\
+\frac{\partial}{\partial t}\left\{ n\varepsilon\left(1+\frac{u^{2}}{2c^{2}}\right)+p\frac{u^{2}}{2c^{2}}\right\} =0\,.\label{eq:20}\end{eqnarray}

\noindent It is important to emphazise at this point that the terms
that correspond to the mechanical energy in the total energy balance
arise from the expression

\begin{equation}
n\varepsilon\frac{u^{2}}{2c^{2}}\sim n\frac{mu^{2}}{2}\,,\label{eq:21}\end{equation}

\noindent which would not be present if the second order terms in
$\mathcal{G}^{44}$ are considered, as mentioned above. Thus, the
energy balance is

\begin{eqnarray}
\frac{\partial}{\partial t}\left(\frac{3}{2}nkT+nm\frac{u^{2}}{2}\right)+\phantom{gffgfdhjfhjj}\nonumber \\
+\left[\left(\frac{3}{2}nkT+p+nm\frac{u^{2}}{2}\right)u^{\ell}\right]_{,\ell}=0\label{eq:22}\end{eqnarray}

\noindent or \begin{equation}
\frac{\partial T}{\partial t}+\vec{u}\cdot\nabla T+\frac{2}{3}T\nabla\cdot\vec{u}=0\,,\label{eq:23}\end{equation}

\noindent which is the usual expression for the temperature evolution
in Euler's regime.

\section{Discussion and Final Remarks}

In the previus sections we have archived a 3+1 representation for
Euler hydrodynamics. The link between fluxes in the comoving and laboratoy
frame is given by the pseudo-galilean transformation given in Eq.
(\ref{eq:9}). This matrix is obtained as the non-relativistic limit
of the Lorentz transformation. Notice that the $\mathcal{G}^{\ell4}$
components are odd in $u/c$ and, as can be easily seen, the next
correction is third order in this quantity. Thus, the proposed transformation
is the non-relativistic limit of the Lorentz transformation to second
order in $u/c$. As pointed out before, by keeping these second order
terms, the formalism is capable of producing the energy balance equation
from the energy-momentum tensor conservation law. Notice that, under
this transformation, the position vector in the laboratory frame is
given by\begin{equation}
\bar{x}^{\ell}=x^{\ell}+u^{\ell}t\label{yy}\end{equation}
 \begin{equation}
\bar{x}^{4}=ct\left(1+\frac{u^{2}}{2c^{2}}\right)+\frac{u^{\ell}x_{\ell}}{c}\label{y}\end{equation}
from which we could characterize the transformation as purely galilean
in space and semi-relativistic in time. This suggests that the usual
galilean transformation is not the correct aproximation for the non-relativistic
limit but an {}``ultra non-relativistic limit''.

Besides providing an alternative method for obtaining the Euler equations
for the non-relativistic fluid, the formalism proposed serves as a
benchmark for the relativistic case. In special relativity, not only
the transformation law for the velocities is not a linear relation
but also a relativistic local equilibrium distribution function must
be considered. In that case the integrals involved in calculating
the averages are not trivial to perform and the physical meaning can
be lost in the lengthy and complicated algebra. However, the calculation
of the fluxes in the comoving frame features no major complications.
Thus, the formalism here developed seems a viable alternative for
studying the dynamics of relativistic fluids in equilibrium.

\end{document}